\documentclass[onecolumn,showpacs,preprintnumbers,amsmath,amssymb,bbold]{revtex4}
\usepackage{epsfig,amsopn}
\usepackage{graphicx}
\usepackage{epstopdf}
\usepackage{sidecap}
\usepackage{amsmath,amssymb}
\usepackage{amsthm}
\usepackage{enumerate}
\usepackage{bbold}
\usepackage{color}

\newcommand\bea{\begin{eqnarray}}
\newcommand\eea{\end{eqnarray}}
\newcommand\beq{\begin{equation}}  
\newcommand\eeq{\end{equation}}

\newcommand{\non}{\nonumber}  
\newcommand{\prx}{Phys. Rev. {\bf X}}

\begin{document}

\title{Weyl semi-metals : a short review}
\author{Sumathi Rao}
\affiliation{Harish-Chandra Research Institute, Chhatnag Road, 
Jhusi, Allahabad 211 019, India }

%
 \begin{abstract}
 
 We begin this review with an introduction and a discussion of Weyl fermions as emergent particles in condensed matter systems,
and explain how high energy phenomena like the chiral anomaly can be seen in low energy  experiments.
We then explain the current interest in the field due to the recent discovery of real materials which behave like Weyl semi-metals.
We then describe a simple lattice model of a topological insulator, which can be turned into a Weyl semi-metal
on breaking either time-reversal or inversion symmetry, and show how flat bands or Fermi arcs develop. Finally, we describe some 
new phenomena which occur due to the chiral nature of the Weyl nodes and end with possible future prospects in the field, both 
in theory and experiment.

 \end{abstract} 

\maketitle

\section{Introduction}

In recent times, the Landau paradigm of classification of states through the principle of spontaneous symmetry breaking and local 
order parameters has been superseded by the classification through topology or global properties of the many-body ground state
wave-function of the system\cite{classification}. These topological phases have some property to which an integer can be assigned, which is robust
and depends only on global properties. They cannot be destroyed by local perturbations such as disorder and scattering, as long
as the bulk gap is not closed, although they can have surface metallic states - in fact, surface metallic states are what distinguishes them
from ordinary insulating phases.

The first, most famous example of a topological phase is the integer quantum Hall effect\cite{iqhe}.  This was a system of electrons moving in
two dimensions, with a strong perpendicular magnetic field (time-reversal broken system). This led to Landau levels and most importantly,
to the quantisation of the transverse conductance with remarkable accuracy.  It was soon realised that 
this effect  had a topological explanation\cite{tknn} - the physically measured transverse
current was related to a topological invariant,  the first Chern number, which is just the integral of the Berry curvature over the Brillouin zone.
The topological phase on each plateau was protected by a bulk gap, and the current was carried by metallic  surface or edge states.

However, for a long time, this did not have much impact on material science, because quantum Hall effect occurred
at low temperatures and at very high magnetic fields.  There was  a lattice model proposed by Haldane\cite{haldane1}
which suggested that materials could have topologically nontrivial  band structures, characterised by non-zero Chern numbers,
even without an external magnetic field.  But it is only 
in the last decade or so  that it was realised that this could occur in realistic materials.  Consequently,
there has been an explosion of work on topological materials\cite{ti}. The hallmark of these materials was that they were
insulating in the bulk, but they could conduct electricity through metallic surface states,
without magnetic field or breaking time-reversal invariance. Hence, 
unlike the quantum Hall phases, these materials had to have a pair of counter-propagating edge states due to time-reversal invariance.
Also, unlike the quantum Hall phases, they could occur in three dimensions as well as in planar systems.
Furthermore, in these systems, the edge or surface states had  relativistic dispersions and hence, the physics of Dirac fermions
became relevant in these materials.

But  more recently, it has been realised that gaplessness is not an essential ingredient for topological protection.
Band topology can be defined, even if the gap closes at some points in the Brillouin zone.  A particular example of 
such a phase is the Weyl semi-metal phase\cite{murakami}, which was first predicted in the pyrochlore iridates\cite{wan}.
It is a new  state of matter, whose low energy excitations are Weyl fermions.  Unlike the topological insulators, 
these materials have   gapless states in the bulk as well as the boundary.

\section{Emergent Weyl fermions in condensed matter systems}

The idea that  particles with relativistic dispersion can be quasiparticles in condensed matter systems
is not new and goes back to the early days of the study of Luttinger liquids\cite{onedfermions}, where it was realised that
close to the Fermi energy, the dispersion could be linearised and obeyed relativistic energy-momentum relations.

More recently, relativistic Dirac fermions came into prominence with the discovery of graphene\cite{graphene} in 2+1 dimensions.
Graphene is a single sheet of carbon atoms arranged in a honeycomb lattice. Electrons moving around the carbon atoms
interact with the periodic potential of the honeycomb lattice to give rise to a Fermi surface with six double cones
where the valence and conductance bands touch each other. (Out of these six, only two are independent).  Close to
these nodes, the dispersion  is  linear and the Hamiltonian is  given by the massless relativistic Dirac equation,
with the speed of light replaced by the Fermi velocity and the spin replaced by the pseudo-spin of the sub-lattices -
\beq
H =  v_F(\sigma_x p_x +\sigma_y p_y)~.
\eeq
The wave-functions are   two component spinors  with the spinor index refering to the sublattice index and real
spin being  an additional quantum number. $v_F$ is  the Fermi velocity in the solid, which replaces the speed of light
in relativistic systems.  The low energy excitations about the nodes  are gapless (graphene is metallic).
However, in principle, they can be gapped out by perturbations (mass terms) proportional to $\sigma_z$.
 The stability of graphene comes from the extra symmetries under time-reversal and 
spatial inversion\cite{guinea}, which  enforces the vanishing of terms proportional to $\sigma_z$ and hence
implies that no gap is induced, as long as perturbations do not break time-reversal and inversion symmetry.

A possible generalisation of this to 3+1 dimensions is to write down the massless 3+1 Dirac Hamiltonian
given by 
\beq
H = \begin{pmatrix} 0 &v_F {\vec \sigma}\cdot {\bf p} \\ v_F{\vec \sigma}\cdot {\bf p}  & 0\end{pmatrix}
\eeq
This  is equivalent to having two doubly degenerate Dirac cones at the same point in momentum space. This is what has been
dubbed as a Dirac semi-metal\cite{dmetals}.  This semi-metal can also  be gapped out easily by adding  mass terms proportional
to a diagonal $4\times 4$ matrix ($I_{2\times 2}, - I_{2\times 2}$).  However, the Dirac points can be protected by crystal symmetries, and
provided that the perturbations or imperfections which exist do not break this symmetry, semi-metallic phases can be observed.

However, it is possible to split the degeneracy of the Dirac node in momentum space or energy space 
by breaking either time-reversal
or inversion symmetry. This gives rise to two Weyl nodes with opposite chiralities (or helicities), 
whose dispersion is given by the massless Weyl Hamiltonian. 
The basic idea\cite{murakami}  behind the existence and stability  of these  Weyl nodes is very simple. 
Close to the degeneracy points where two 
non-degenerate bands
touch each other and the energy is cone-like, the electronic  excitations are described by the Weyl equation, simply because the effective
Hamiltonian at that point has to have two dimensions, and besides the identity, there are  only  three 
anti-commuting $\sigma$ matrices. 
More explicitly,  the low energy Hamiltonian for a two band model is 
given by
\beq
H = a({\bf k}) + \sigma \cdot {\bf b} ({\bf k})~.
\eeq
where $\sigma = (\sigma_x,\sigma_y,\sigma_z)$ are the three Pauli matrices.
We can now expand  the Hamiltonian around the touching or gapless points (${\bf k} = {\bf k}_0$) in the Brillouin zone to get
\beq
H({\bf k} \simeq {\bf k}_0) \simeq {\rm const} + \sigma \cdot {\bf b} ({\bf k}_0) + \sigma \cdot { \frac {\partial b_i} {\partial k_j}}_{({\bf k} =
{\bf k}_0)}({\bf k} - {\bf k}_0)~.
\label{anisoweyl}
\eeq
This is precisely the Weyl Hamiltonian (though offset and anisotropic) and the wave-functions  are two 
component Weyl fermions with positive or negative chirality.  
These band-touching points or nodes can 
move around in the momentum space, by perturbations that change the Hamiltonian slightly,
but the only way for them to gap out and disappear is if they
meet another node with the opposite chirality.  This is very different from what happens in graphene.
In fact, a straighforward  generalisation of the graphene Hamiltonian to 3+1 dimensions  can also be made  simply by adding
a term proportional to $\sigma_z$ to obtain
\beq
H = \sigma_x p_x +\sigma_y p_y +\sigma_z p_z
\eeq
which is just the isotropic form of the model given in Eq.\ref{anisoweyl} and is the familiar Weyl Hamiltonian in relativistic systems.
Here, there is no fourth anti-commuting matrix to add  a mass term. Perturbations can only move the nodes,
but cannot gap them out.
Hence, we have gapless fermions with a single chirality at each band touching point.

Surprisingly, it was found, initially through detailed calculations\cite{wan,burkovbalents,xu},  that there are condensed matter
materials, whose band structure shows nodes or points around which the excitations are Weyl fermions.
These materials were dubbed Weyl semi-metals (WSM)\cite{wsmreviews}.   They also have very unusual surface states called
Fermi arcs, which terminate at the location of the bulk Weyl nodes, which we will discuss in detail later.

As was already mentioned, in a system with both inversion symmetry and time-reversal symmetry intact, one always gets Dirac nodes -
both left and right handed chiral nodes, at the same point, in which case it is always possible to add a mass term using the 
fourth anti-commuting four component $\gamma$ matrix and gap it out. 
So to get Weyl nodes in the condensed matter system,
one needs to break either inversion symmetry or time-reversal symmetry.

The stability of the Weyl nodes can also be connected to a topological quantum number, which is the conservation of the total
charge of the Weyl nodes. As we said above, a  single Weyl node cannot disappear by itself. It has to annihilate with a Weyl node of the opposite charge. This also means that for the Weyl nodes to have well-defined coordinates, momentum has to be a good quantum number, so translational invariance has to be unbroken. So disorder has to be fairly weak.

Also, Weyl nodes can be thought of as monopoles in momentum space. The charge of the Weyl node can be related to  the
quantised Berry flux, which can be computed from the momentum space wave-functions of the filled states near the Weyl node.
It turns out to be $\pm 2\pi\kappa$ depending on the chirality of the node (positive or negative) and the `monopole\rq{} charge $\kappa$\cite{murakami}. 
The topological stability of the Weyl node is thus related to the Gauss  law, which keeps the total flux inside a given surface
invariant. The Weyl node can only vanish when it annihilates with another Weyl node with opposite chirality.
Also, since the net charge of all the Weyl nodes inside a Brilloiuin zone has to be zero, the Weyl nodes
always come in pairs and the minimum number of Weyl nodes that one can have in any model is two\cite{nn}.

\section{The `high energy' connection}

The distance between high energy physics and condensed matter physics has been decreasing in recent years,
with the so-called (AdS-CFT) Anti-de Sitter- Conformal Field theory (or even AdS-CMT for condensed matter theory) 
connection\cite{sachdev}. But in the field of Dirac materials, the connection is much more direct and  the idea that 
quasiparticles can be relativistic fermions is quite old\cite{onedfermions}.  Moreover,  with the discovery of topological materials, 
there has been  a possibility of seeing even Majorana modes (not exactly Majorana fermions, but related) and now
Weyl  fermions in low energy condensed matter settings. We will discuss the discovery of these Weyl fermions in the next section,
but in this section, we will see how phenomena such as the chiral anomaly related to Weyl fermions  gets translated in the
condensed matter context.

The basic idea of the chiral anomaly is that the conservation laws $\partial_\mu j^\mu=0$ and $\partial_\mu j^\mu_5=0$ 
cannot be simultaneously satisifed. So if we take the conservation of current to be sacrosanct, the current of the left
and right chirality fermions cannot be individually conserved. The Adler-Bell-Jackiw anomaly equation\cite{ABJ}
\beq
\partial_\mu j^\mu_5= F F^*/8\pi^2
\eeq
thus leads to breaking of the chiral symmetry by the anomaly. In the Weyl semi-metal context, these equations
can be rewritten as 
\bea
&\frac{\partial}{\partial t} (n_R+n_L) = 0, \implies \frac{\partial}{\partial t} (n_R - n_L) \ne 0, \nonumber \\
&\frac{\partial}{\partial t} (n_R - n_L) =\pm \frac{e^2}{h^2} {\bf E}\cdot {\bf B},
\eea
where $n_R$ and $n_L$ denote the number of fermions at the right and left chirality Weyl nodes.
This implies that by applying parallel electric and magnetic fields, we can change the difference in their numbers,
or if we have really have an isolated Weyl node, we can change the number of particles. This is essentially a quantum mechanical effect
because the path integral for Weyl fermions coupled to an electromagnetic field is not invariant under independent gauge transformations
of the left and right chiral fields. So single chirality fermions implies charge non-conservation.
However, in any real system, one always has total particle number conservation. Hence, Weyl nodes in any crystal always 
have to come in pairs. Due to the anomaly term, we  can get  charge pumping between the nodes  - 
$n_R$ decreases and $n_L$ increases or vice-versa, at a rate given by the anomaly.
But the total number  of particles remains the same.
This is called the Nielsen-Ninomiya theorem\cite{nn}~.

More formally, by coupling the
Weyl  fermions to an external electromagnetic field  and computing the effective action by integrating out the fermions, it can be shown that the time-reversal
or parity breaking parameter in the model leads to an effective $\theta$ term in the  action given by
\beq
S_\theta  = \frac{ e^2}{32\pi^2} \int d^4x \theta(x) \epsilon^{\mu\nu\alpha\beta}F_{\mu\nu} F_{\alpha\beta}, 
\eeq
(we are using 4 vector notation here and $\mu$ runs from 0 to 3)
where the axion field $\theta(x) = 2b_\mu x^\mu = 2{\bf b}\cdot {\bf r} - 2 b_0 t$ and $b_\mu = b_0,b_i$ are parameters that break the parity
 ($b_0$) and time-reversal
($b_i$). The current can be computed from the effective action and is found to be
\bea
j^\nu  &=& \frac {e^2}{2\pi^2} b_\mu \epsilon^{\mu\nu\alpha\beta}\partial_{\alpha} A_{\beta}, \quad \mu =1,2,3 \nonumber \\
{\rm and} \quad j^\nu  &=& \frac {e^2}{2\pi^2} b_0 \epsilon^{0\nu\alpha\beta}\partial_{\alpha} A_{\beta}~. 
\eea
The first equation above describes the anomalous Hall effect, with the Hall conductance proportional to the
separation between the Weyl nodes
\beq
\sigma_{xy} = e^2/h \times |{\bf b}|~. \label{anomHall}
\eeq
The second equation describes the chiral magnetic effect, which naively seems  to imply equilibrium currents in 
the presence of a magnetic field\cite{zyusinetal,zhou1}.
But the notion of the chiral anomaly  in a condensed matter system is itself somewhat idealised, because unlike in the
relativistic case where chiral symmetric is exact for massless fermions, here, even the relativistic dispersion for the
Weyl fermions and chiral symmetry is only approximately true, close to the Weyl nodes. Hence, using the chiral
anomaly directly to predict effects in the Weyl semi-metal is not a very good idea.  More detailed calculations have shown
that the first result can  be rederived in different ways\cite{yang}, but the second result involving 
 the naive expectation of equilibrium currents  is incorrect in the 
condensed matter context\cite{vazifeh}. As pointed out in Ref.\onlinecite{haldane}, in a Weyl semi-metal, although there is a charge
imbalance at the nodes due to the anomaly, there is also a return path for the charges via the Fermi arcs on the surfaces.
 Hence, one needs to
be careful in applying relativistic ideas directly to the Weyl semi-metal system.

So the basic idea  to look for effects of the anomaly, would be to apply parallel electric and magnetic fields to the system and
see whether  there are explicit effects due to the charge imbalance between the two chiral nodes in the condensed matter system itself.
One would expect the results to be 
highly anisotropic,  because the effect depends on the angle between the applied electric field and  the  magnetic field,  being
maximum when ${\bf E} \parallel {\bf B}$ and minimum (vanishing) when  ${\bf E} \perp {\bf B}$. In fact,  there have been
several works \cite{burkov,gorbar} which have explicitly computed the physical effects of the chiral  anomaly in Weyl semi-metals.
They have shown that  the chiral anomaly actually  leads to  large negative longitudinal magneto-resistance
at weak fields, as expected from the anomaly equation. More recently, there has been work\cite{goswami1} which has
shown that the chiral anomaly generically leads to longitudinal magneto-resistance (LMR) in three dimensional chiral metals
(not necessarily Weyl semi-metals and not necessarily even Dirac materials). In a Weyl semi-metal, the sign of the LMR depends 
on the kind impurities in the sample, and generically, with both ionic and neutral impurities,
the LMR becomes negative initially and then becomes positive as a function
of the magnetic field. 

Other effects of the anomaly such as current induced by strain fields\cite{zhou1}, anisotropic non-local voltage
drops\cite{vish} and optical absorption\cite{ashby} due to the charge imbalance between the two nodes have also been studied. 
For inversion symmetry breaking semi-metals, the chiral
nodes are at two different energies, which leads to a chiral chemical potential between the nodes. This can give
rise to measurable optical signals\cite{goswami}. Other effects include unusual plasmon modes\cite{zhou} in doped WSM.
All these possibilities have spurred experimental activity in this field and in the last few months, there has been definite evidence for
Weyl fermions in several materials.

\section{Current excitement due to `discovery\rq{} of Weyl fermions}

The initial proposals for Weyl semi-metals included pyrochlore iridates\cite{wan}, $HgCr_2Se_4$\cite{xu} and 
topological  insulator-normal insulator heterostructures\cite{burkovbalents}, but there is no
experimental evidence for them yet. Negative magnetoresistance was  observed experimentally\cite{kim}
when magnetic field was applied to   $BiSb$  tuned to its critical point (the Dirac point) between normal
and topological insulator, but that by itself was not convincing enough to confirm the existence of Weyl fermions,
particularly since it was realised  that negative magneto-resistance is generic in three dimensional metals.
However, in the last few months, there has been a lot of excitement\cite{manypapers} in the field, ever since  the publication of 
 direct evidence for not only the Weyl nodes, but
also  the Fermi arcs in the non-centrosymmetric  material $TaAs$
using ARPES (angle resolved photoemission spectroscopy) techniques,
by the Princeton group\cite{hasan} and a  group from the Chinese Academy of Sciences\cite{lv}.
They were able to show co-propagating Fermi arcs terminating at the Weyl nodes with non-zero chiral charges
of $\pm 2$.
At the same time, Weyl nodes were also seen in a photonic  crystal, which was `made to order'
to reproduce a desired band structure\cite{photonic}.   Both these cases involved inversion symmetry breaking.
In $TaAs$, the crystal structure itself lacked inversion symmetry and in the photonic crystal, inversion symmetry breaking 
was explicitly incorporated. The photonic crystal was made such that it realises 4 Weyl nodes (the minimum for
inversion symmetry breaking Weyl semi-metals), whereas $TaAs$ has 24 Weyl nodes.
Fermi arcs have also been observed in Dirac semi-metals, such as $Cd_3As_2$\cite{neupane} and $Na_3Bi$\cite{liu}.
Other candidate materials include an inversion symmetry breaking stochiometric compound $SrSi_2$\cite{huang}
which forms a WSM without spin-orbit coupling and an exotic state with quadratic  dispersion when spin-orbit coupling is
included, but this has not yet been confirmed. For a recent review of topological semi-metals predicted from first principle
calculations, see [\onlinecite{hongming}]~.

\section{Explicit lattice model and phase diagram}

\begin{figure}
  \includegraphics[width=0.35\textwidth]{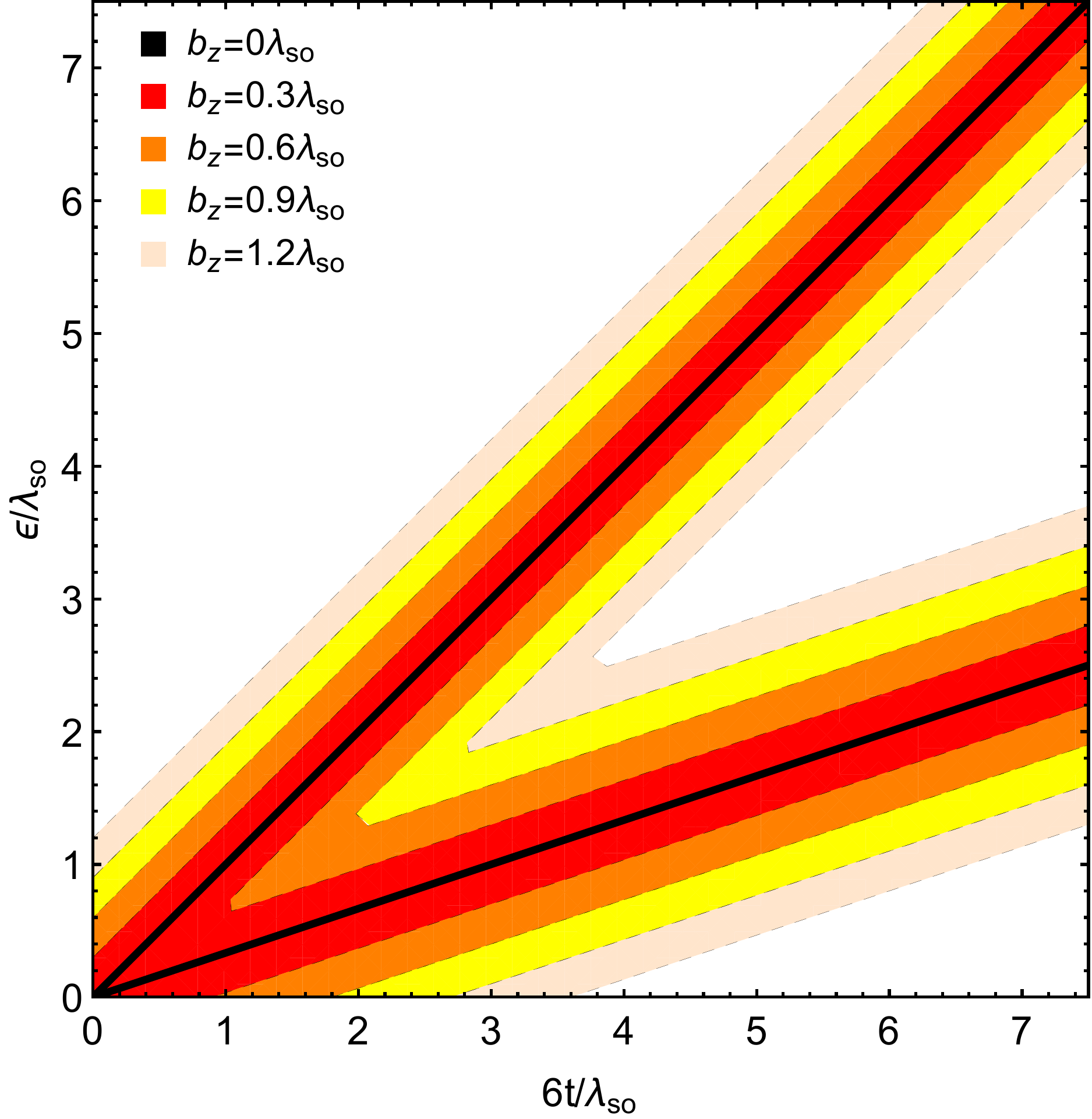}
  \includegraphics[width=0.35\textwidth]{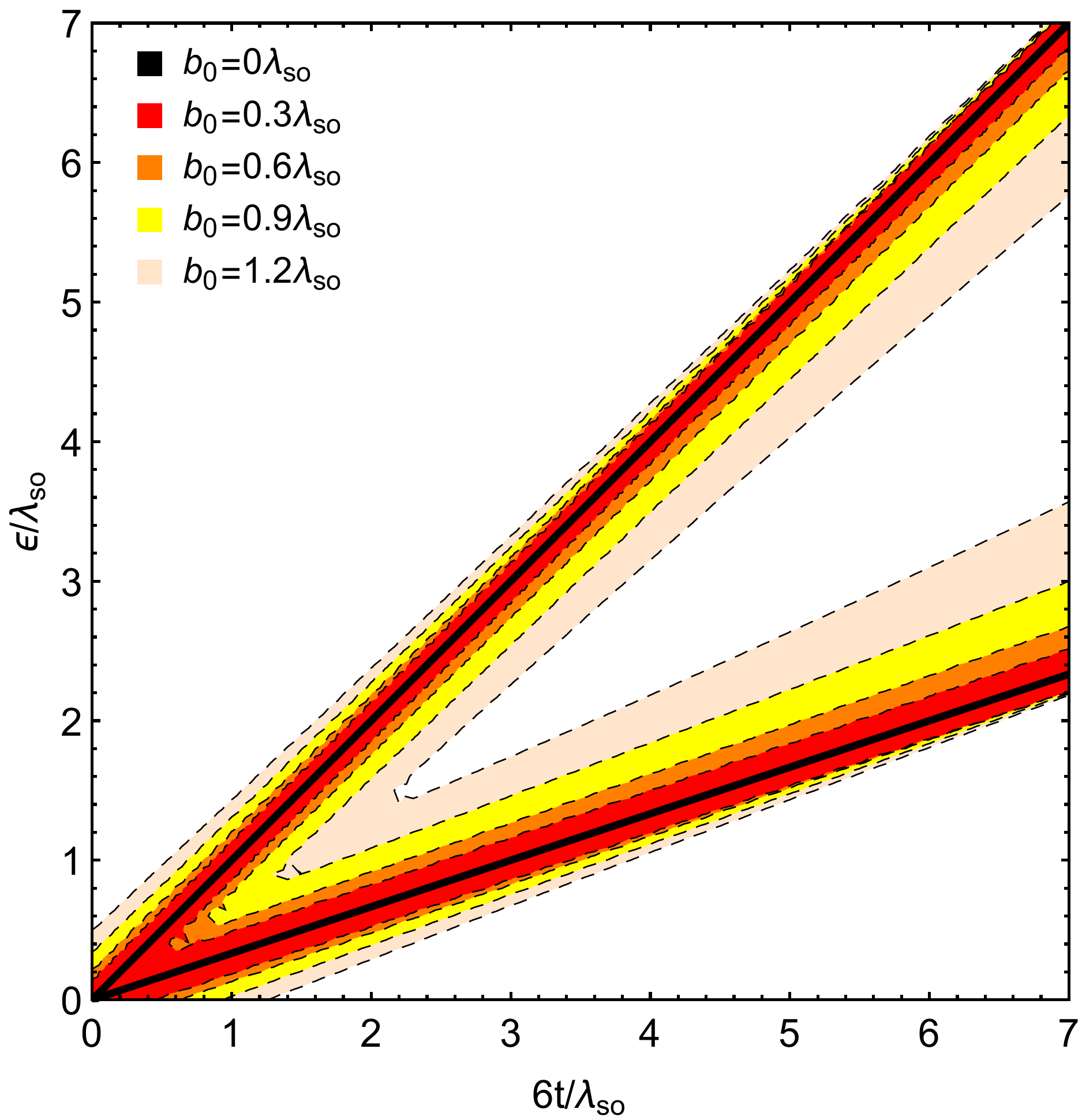}
  \caption{(Color online) Typical phase diagrams of our model system. The Weyl semi-metal (WSM) phase appears at the strong topological insulator (STI)/ normal insulator (NI) ($\epsilon = 6t$) and strong topological insulator (STI)/ weak topological insulator (WTI)  ($\epsilon = 2t$) boundaries (shown by black lines) with broken time reversal ($b_z$)/ parity ($b_0$)  perturbations. The WSM phase extends with increasing perturbations in the filled regions. Parameters used here are $\lambda_{z} = \lambda_{\text{SO}}$.}
   \label{fig:phasediagram1}
\end{figure}

We will now study some features of the WSM in detail\cite{proximity},  using  a simple tight-binding four-band lattice model \cite{vazifeh,delplace} for the topological insulator (TI) in three dimensions (3DTI), 
whose phases  can describe strong and weak topological insulators, Weyl semi-metals and ordinary insulators depending on the values of the various 
parameters of the model. Many 3DTI materials, including the family belonging to the  $\text{Bi}_2\text{Se}_3$ class,  have an effective
description in terms of the  Hamiltonian given by 
$H_0 =  H_{\text{C}}+ H_{\text{SO}}$ \cite{vazifeh} with 
\begin{align}\label{eq:h0}
 H_{\text{C}}&=\epsilon \sum_{j} \psi^{\dagger}_{j}\tau_x\psi_{j} - t\sum_{\langle {ij} \rangle}\psi^{\dagger}_{i}\tau_x\psi_{ j} + ~\text{h.c.} \nonumber \\
  {\rm and} ~~~ H_{\text{SO}} &= i\lambda_{\text{SO}}\sum_{j} \psi^{\dagger}_{ j}\tau_z\left( \sigma_x\psi_{j+\hat{y}} - \sigma_y\psi_{j+\hat{x}} \right) \nonumber \\
 &+i\lambda_z\sum_{j}\psi^{\dagger}_{j}\tau_y\psi_{j+\hat{z}} +  ~\text{h.c.}
\end{align}
Here $z$ is the direction of growth of the crystal and  $\bf{\sigma}$ and $\bf{\tau}$ denote Pauli matrices in spin and orbital(parity)  space respectively.
$\epsilon$ and $t$ denote the standard on-site and nearest neighbour hopping amplitudes.
 $\lambda_{\mathrm{SO}}$ and $\lambda_{z}$ are the (possibly anisotropic) spin-orbit (SO) interaction strengths 
 in the $x$-$y$ plane and in the $z$ direction respectively. $i,j$ refer to site indices in all three dimensions.
 The  topological invariants for the  3DTI, $\nu_{\mu} = (\nu_0; \nu_1, \nu_2, \nu_3)$ can be easily obtained  using
parity invariance\cite{fukane} and are  given by\cite{proximity}
\begin{align}
 (-1)^{\nu_0} &= \text{sgn}\left[ (\epsilon - 6t) (\epsilon + 6t) (\epsilon - 2t)^3 (\epsilon + 2t)^3 \right], \non \\
(-1)^{\nu_i} &= \text{sgn}\left[ (\epsilon + 6t) (\epsilon - 2t) (\epsilon + 2t)^2 \right], \nonumber
\end{align}
for  $i = 1, 2, 3$. Depending on the values of the invariants,  we have the following phases:
\begin{align}  \begin{array}{ccc}
 |\epsilon| > |6t| & \nu_{\mu} = (0;0,0,0) & \text{Ordinary Insulator} \\
 |6t| > |\epsilon| > |2t| & \nu_{0} =1 & \text{Strong Topological Insulator (STI)} \\
 |2t| > |\epsilon| > 0 & \nu_{0} = (0;1,1,1) & \text{Weak Topological Insulator (WTI)}\end{array} 
\end{align}

The Weyl semi-metal (WSM) phases arise close to the boundaries of the topological phase transitions, at $\epsilon \approx \pm 6t, \pm2t$ 
when either  parity (inversion) or  time reversal (TR) symmetry 
or both  are broken.  In other words, the Hamiltonian for the WSM is given by 
 $H_W = H_0 + H_{\text {E}}$  where 
\begin{equation}\label{eq:he}
 H_{\text{E}} = \sum_{j} \psi^{\dagger}_{j} \left( b_0\tau_y\sigma_z - b_x\tau_x\sigma_x + b_y\tau_x\sigma_y + b_z\sigma_z\right)\psi_{j} ~.
\end{equation}
Here $b_0$ and ${\bf b}$  are parameters that break inversion and TR symmetry
respectively and separate the Dirac point of the TI into two Weyl points, with the 
separation being in energy and in momentum  space respectively\cite{vazifeh}. 
An arbitrary hermitean term can be added  to this Hamiltonian, whose effect will be to cause the Weyl nodes
to move around. However, for small perturbations, it cannot remove the nodes. The nodes can only vanish if the perturbation
is large enough to bring both the Weyl nodes together.
The phase diagram of the
different phases in this model is given in Fig.~\ref{fig:phasediagram1}\cite{proximity}. 

\begin{figure}[ht]
\includegraphics[width=0.48\textwidth]{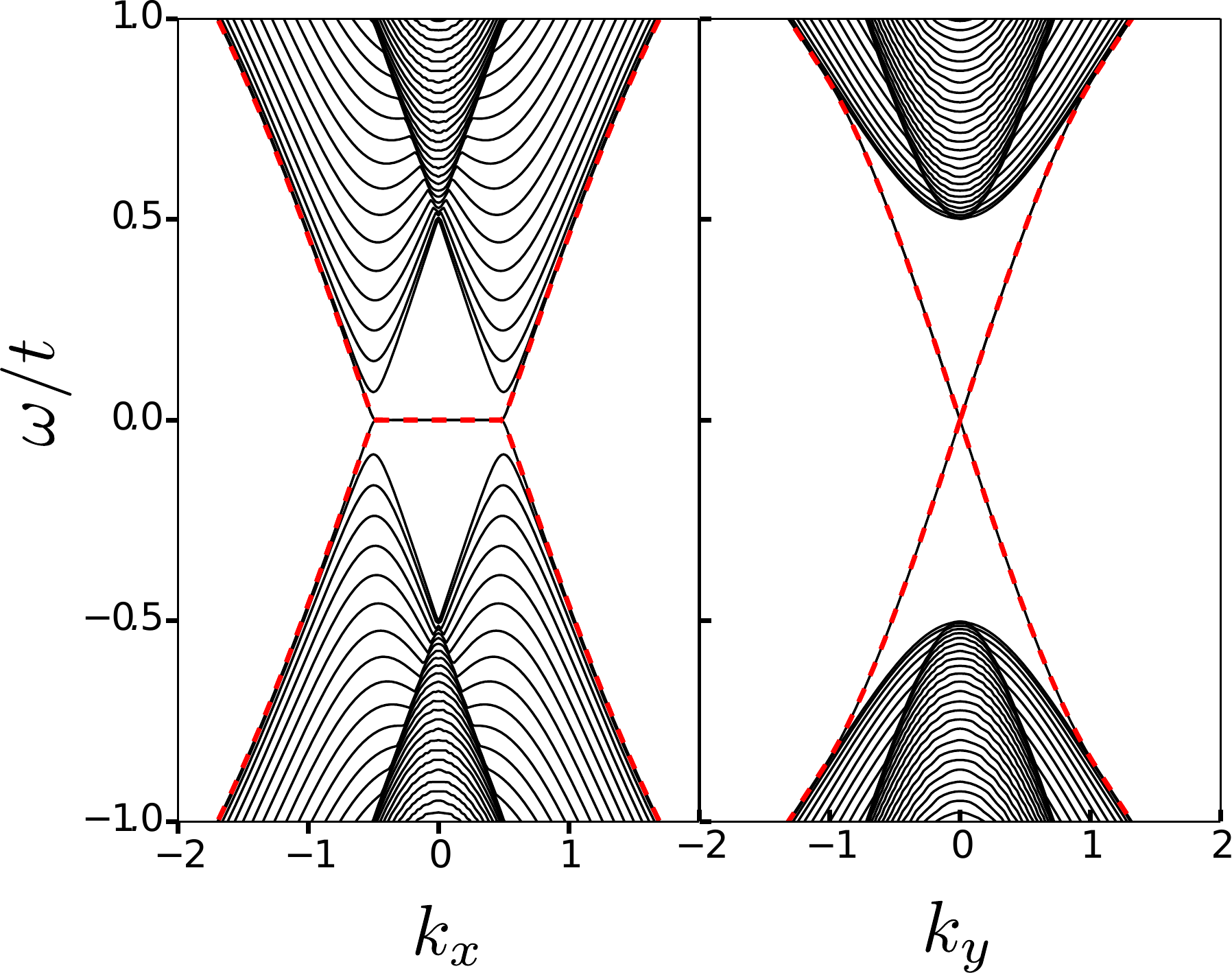}
\caption{(Color online) The  dispersion for a WSM with 2 Weyl nodes at $k_0=\pm b_x /\lambda_z$ is shown  along  $k_x$ and  $k_y$.
The parameters used are ${\bf b}=(0.50t,0,0), \lambda_{SO}=\lambda_z=0.50t$. The dashed (red) lines denote the surface band.}
\label{dispersion}
\end{figure}

\begin{figure}[ht]
\includegraphics[width=0.48\textwidth]{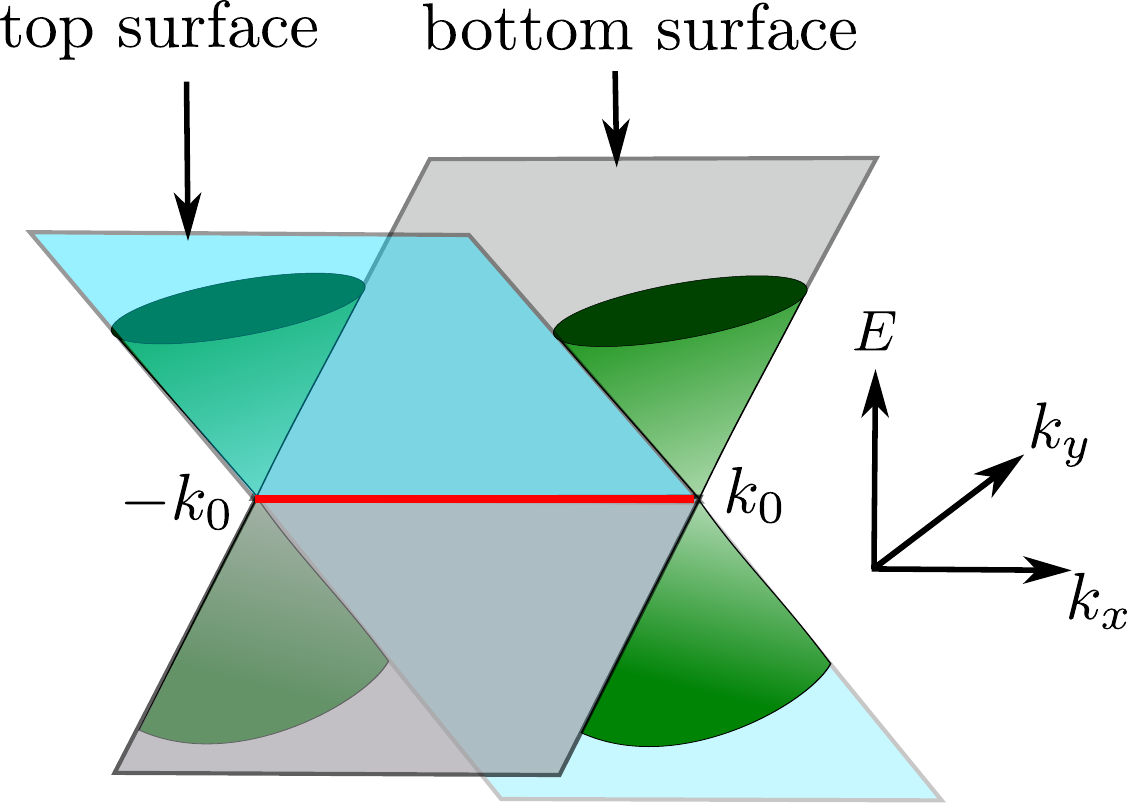}
\caption{(Color online) The  dispersion for a WSM with 2 Weyl nodes at $\pm k_0$. The blue and grey surfaces
show the dispersions of the surface states on the top and bottom surfaces respectively. The cones represent the bulk states and the picture
shows how the surface states merge into the bulk  }
\label{Fermiarcmtm}
\end{figure}

The band structure of the WSM is given in Fig.~\ref{dispersion}\cite{proximity}. As can be seen from there, the  cones which describe the Fermi surface around each of the nodes are independent for small enough doping (small values of the chemical potential or Fermi energy $E_F$). But for larger
values of the chemical potential, they merge into a single Fermi surface. For sufficiently large  system size, the dispersion is flat between the 
two Weyl nodes along the $k_x$ direction,  and is linear
along $k_y$ (for $k_x$ values between the Weyl nodes).   These are the surface states (in momentum space) and are shown in Figs.~\ref{dispersion}
and ~\ref{Fermiarcmtm}.

\section{Characteristics of surface states and Fermi arcs}

The existence of surface states for the TI is well-established. They are isolated from the bulk states, since the bulk is
gapped, and they are spin-momentum locked - i.e., the spin is tied to the direction of motion, and opposite spins 
travel in opposite directions. This has been explicitly seen in many experiments\cite{expTIsurface} using scanning tunneling
microscopy.  Surface states also exist for the WSM. This is more unexpected because one does not expect well-separated surface
states in a gapless model, since they are expected to merge with the bulk states.
So the first question that one has to answer is why surface states exist at all. It is 
because the system (to a good approximation)  is translationally invariant. Hence,  momentum is a good quantum number 
and it is possible
to have surface states at momenta where there are no bulk states - ie., between the Weyl nodes.

It is also found that the surface states at the Fermi energy in WSM form an arc and not a full Fermi surface.
These Fermi arcs end at the projection of the Weyl nodes onto the surface, where the surface mode
is no longer well-defined, because it merges into the bulk modes.
What does this mean? An intuitive picture \cite{wsmreviews} to understand Fermi arcs is to start with a very thin sample of Weyl semi-metal and slowly increase the
thickness so that the two surfaces are pulled apart. To start with, we have a complete Fermi surface, but as the thickness increases,
complementary parts of the Fermi surface get attached to the two surfaces and they get connected to one another through the
Weyl nodes in the bulk.  A picture of how the surface states at the two opposite surfaces  with opposite chiralities 
get connected through the bulk states
is given in Fig.~\ref{Fermiarcmtm}.  This is essentially the same dispersion that is shown in Fig.~\ref{dispersion} separately
along the $k_x$ axis and $k_y$ axis.

\begin{figure}[ht]
\includegraphics[width=0.48\textwidth]{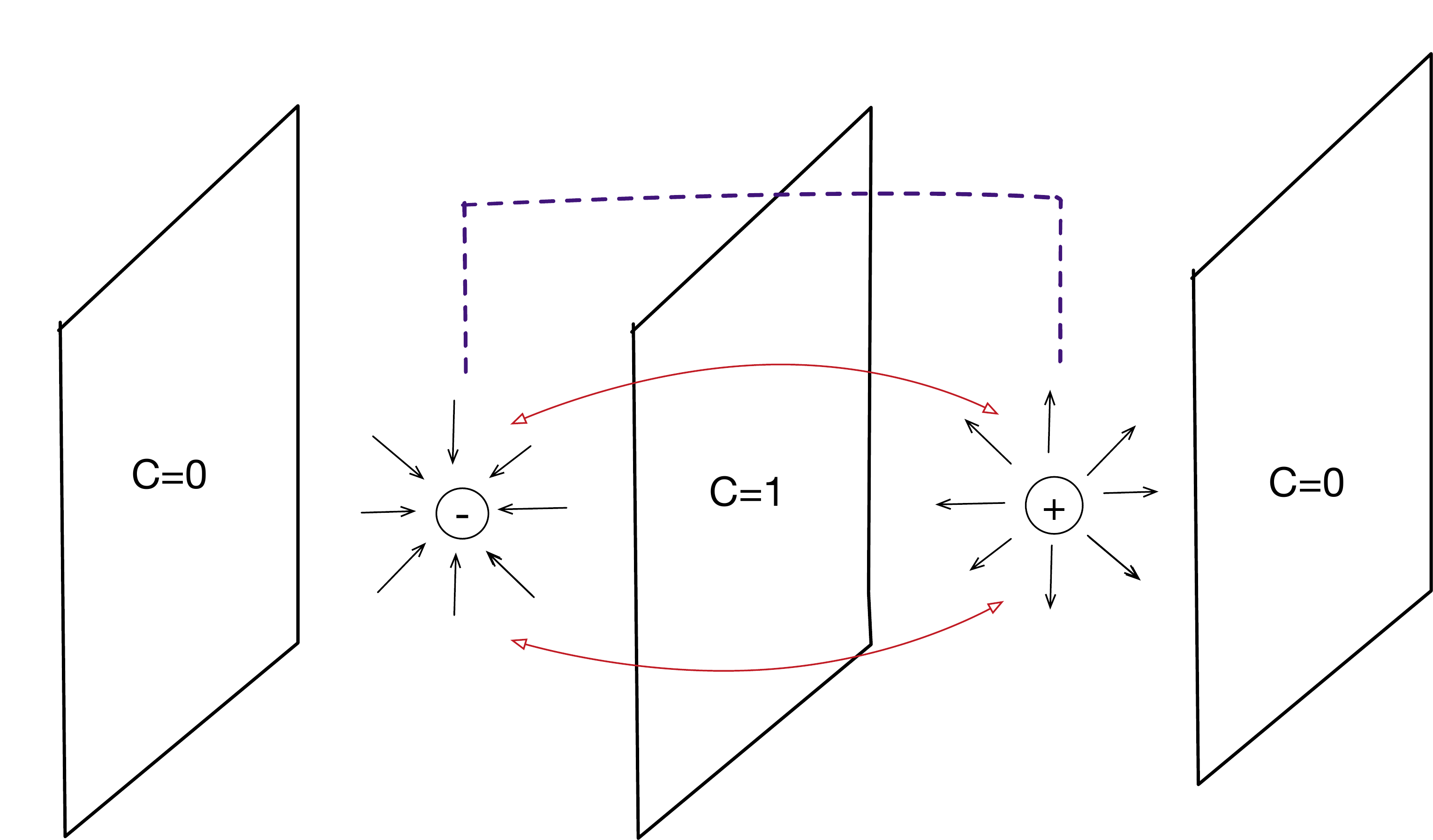}
\caption{(Color online) The  Berry phase monopoles of charge $\pm 1$ at the two Weyl nodes of opposite chirality. The two dimensional planes
within the nodes are quantum Hall planes with non-zero Berry flux passing through it and hence non-zero Chern number. The Fermi arcs
are the edge states of the quantum Hall planes strung together. }
\label{weylmonopole}
\end{figure}

\begin{figure}[ht]
\includegraphics[width=0.48\textwidth]{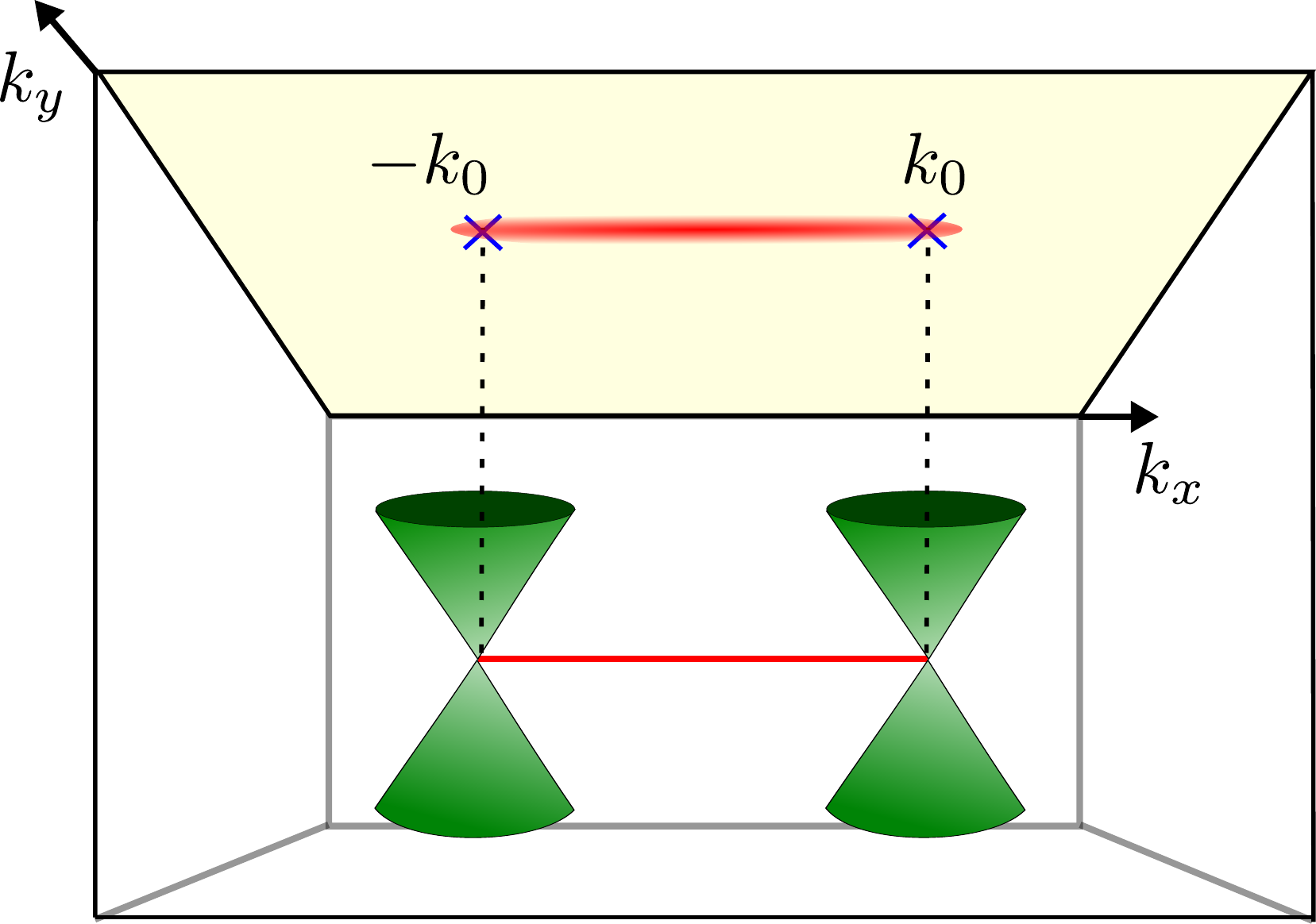}
\caption{(Color online) The  local (surface) density of states (LDOS) for  a WSM with 2 Weyl nodes at $\pm k_0$. 
The red line shows finite value for LDOS, whereas the yellow region denotes zero LDOS. Note that surface states appear
on the  $x$-$y$ and $x$-$z$ planes, but not along the $y$-$z$ planes. }
\label{Fermi-arc}
\end{figure}

We now  study the surface states  for both the TI and WSM phases in the  specific four band (finite-size)  model that we have studied in the previous section  where TR is broken. For strong topological insulators, 
surface states exist on  each surface as mid-gap states\cite{ti}  in the band structure, whereas in weak topological insulators, 
surface states arise only on  particular surfaces depending on the values of $\nu_i$ ($i=1,2,3$)\cite{weakTI}. 
For Weyl semi-metals, in our model, inversion symmetry breaking does not give rise to surface states.
Surface states arise only when TR is broken 
 at the phase boundary between the normal and the topological insulator. 
 For instance, if we choose the TR symmetry to be broken  by  ${\bf{b}} = b_x \hat{x}$, then each
Dirac point yields a pair of Weyl nodes  with a separation of $b_x/\lambda_z \hat{x}$ in momentum space, and we note that
away from the Weyl nodes, there is a gap in the spectrum.
If we put the chemical potential at the energy of the Weyl nodes, we can consider the state obtained by 
filling all the negative energy states. By studying how this state evolves
as a function of the crystal momentum, we define the Berry phase and the Berry flux $\mathcal{B}(k)= \nabla_k \times \mathcal{A}(k)$.
The Weyl nodes here are the sources of this flux $\nabla\mathcal{B}(k)=\pm\delta^3(k-k_\pm)$.

Since the Weyl nodes can be thought of as sources and sinks of Berry phase
monopoles, there is a flux penetrating all the two-dimensional layers between the nodes. All these layers have non-zero Chern number,
whereas two-dimensional planes not between the nodes have zero Chern number. So the two-dimensional planes between the nodes
are like quantum Hall planes and they all exhibit integer quantum Hall effect and have edge states. In other words, for each value of $k_x$  between the 
Weyl nodes, we have a Chern insulator.  This is precisely the reason
that there is an anomalous Hall effect as shown in Eq.\ref{anomHall} proportional to the separation between the Weyl nodes. The Fermi arc is just the line
that one gets by stringing the edge states of the Chern insulator 
 together and clearly these states  only exist between the Weyl nodes. This is illustrated in Fig.~\ref{weylmonopole}~. It also clear from the
 figure that surface states can only exist on the $x$-$y$ and $x$-$z$ planes. No surface states exist on the $y$-$z$ planes since the
 separation between the Weyl nodes is along the $x$-axis.

Now, let us look at the surface states of this model in detail. The first surprise is that there are 
surface states, because there is no gap.   So  if we put the chemical potential at the position of the Weyl nodes, there are not 
only  gapless surface states, but also gapless
bulk states at  the same chemical potential at the nodes.  So we may think that the surface states should hybridise with the bulk states.
But as we mentioned earlier, momentum conservation implies that  it is possible
to have surface states at momenta where there are no bulk states - ie., between the Weyl nodes. So at these values of the momentum,
we will only have surface states and no bulk states - the material will appear like a topological insulator. When we look at the 
the local density of states of the real position-space surface of the three dimensional WSM
 as a function of the momentum,  at the Fermi energy  (tuned to the Weyl nodes), 
we find that they form an arc, (in this particular model, it forms a straight line) 
between the momenta $k_0$ and $-k_0$, the positions of the Weyl nodes, instead of a closed curve.
This is shown in Fig.~\ref{Fermi-arc}. At the Weyl nodes, there is no distinction between the surface states and bulk states, and they
merge into the bulk. But between the Weyl nodes, the surface states  are well-defined. As discussed above, the surface states
are essentially  the edge states of the quantum Hall planes strung together.

There has also been a detailed study\cite{murakami2} of the evolution of the surface states (Fermi arcs) to form a single Dirac cone
as the parameters  of a time-reversal (TR) symmetric  but inversion symmetry broken WSM  are changed to become a TI.

As we mentioned earlier, the excitement in this field has increased many-fold\cite{manypapers} since the actual visualisation of the Fermi
arcs by the Princeton and Chinese Academy of Sciences groups\cite{hasan,lv}~.

\section{New phenomena in Weyl semi-metals}

Weyl semi-metals have bulk conducting states and hence, are metallic. However, there are many ways in which their
behaviour is different from usual metals. The metallic states exist only for some values of momenta and at other
values of momenta, there is a gap to excitations. The excitations at the nodes are chiral and hence their spin is aligned
to the direction of motion.  All of this leads to new transport phenomena which exists only for Weyl semi-metals.
In this section, we will discuss some new results that occur when Weyl semi-metals are placed in proximity with
superconductors. 

\begin{figure}
  \includegraphics[width=0.3\textwidth]{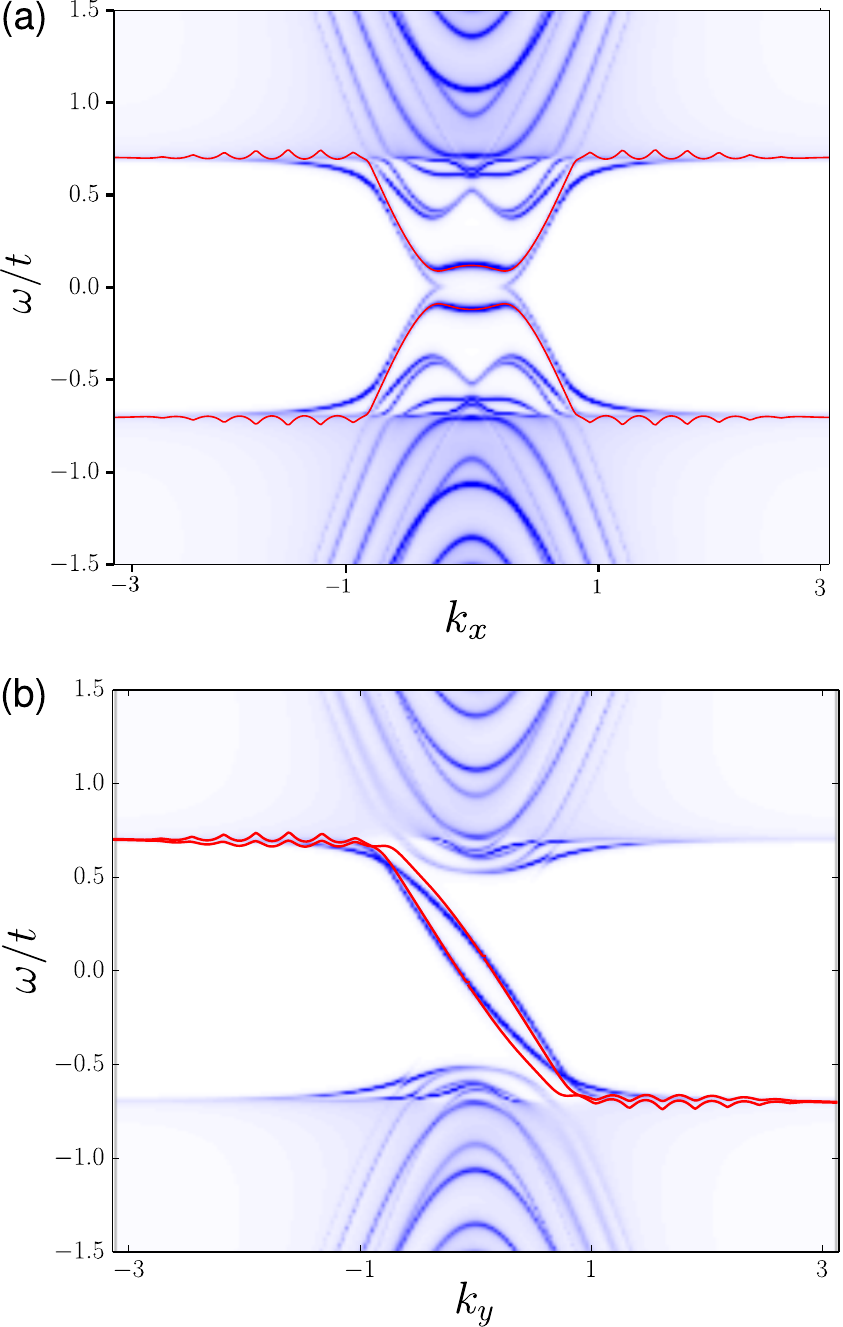}
  \caption{(Color online) The effect of the proximity induced superconductivity in the surface bands of the WSM with momenta (a) $k_x$ and (b) $k_y$ where the Weyl nodes lie along $k_x$. The blue (gray) high density lines are the modified bands in the system with proximity to the superconductor obtained from the LDOS at $z=0$ using a Green\rq{}s function technique,  while the red (darker) solid line is the surface band at $z=0$ via exact diagonalization. The induced gap vanishes at the Weyl nodes for a large enough system size, but the surface band splits. Various parameters used for the LDOS are $\lambda_{\text{SO}}=\lambda_z=0.5 t$, $\Delta=0.7t$, $\lambda_S = 0.9t$, $\epsilon = 6t, ~ {\bf{b}}=(0.5t,0,0)$ and number of sites in $z$ is 20. We have used  $k_y = 0$ for (a) and  $k_x = 0$ for (b).}
   \label{gapfermiarc}
\end{figure}

Motivated by the fact that the introduction of superconductivity via either proximity effect or by
introducing a pairing $\Delta$ term in the bulk leads to a new topological phase in the topological insulator,
we studied\cite{proximity} what happens when a Weyl semi-metal is tunnel coupled to an $s$-wave superconductor.
We started with the model described in Sec.(IV) with two Weyl nodes separated by the time-reversal breaking parameter ${\bf b}$
in the $x$-direction. We took the model to be infinite in the $y$ direction so that $k_y$ is a good quantum number and we took
$z$ to be finite, so that the surface states existed  at the top and bottom in the $z$-direction.
We coupled one of the surfaces of the WSM to an $s$-wave superconductor  and then 
integrated out the superconducting degrees of freedom to obtain  an effective action for the WSM.
We then obtained the local density of states (LDOS) and the induced pairing in the WSM using a Green\rq{}s function
technique as well as exact diagonalisation.
We found that the flat band due to the Fermi arc states split into two when proximity coupled
to the WSM, each carrying half the Chern number of the original surface state, but it does not gap out fully.
The gap remains at the Weyl nodes. This is seen in Fig.\ref{gapfermiarc}\cite{proximity}, and can be understood as follows.
The $s$-wave superconducting correlations couples the electrons at one node of a certain chirality to holes
at the other node, but of the same chirality (because the two nodes have opposite chirality, but electrons and holes
also have opposite chirality). Hence, no gap can open up, because a mass term requires fermions of opposite chirality.

\begin{figure}
  \includegraphics[width=0.45\textwidth]{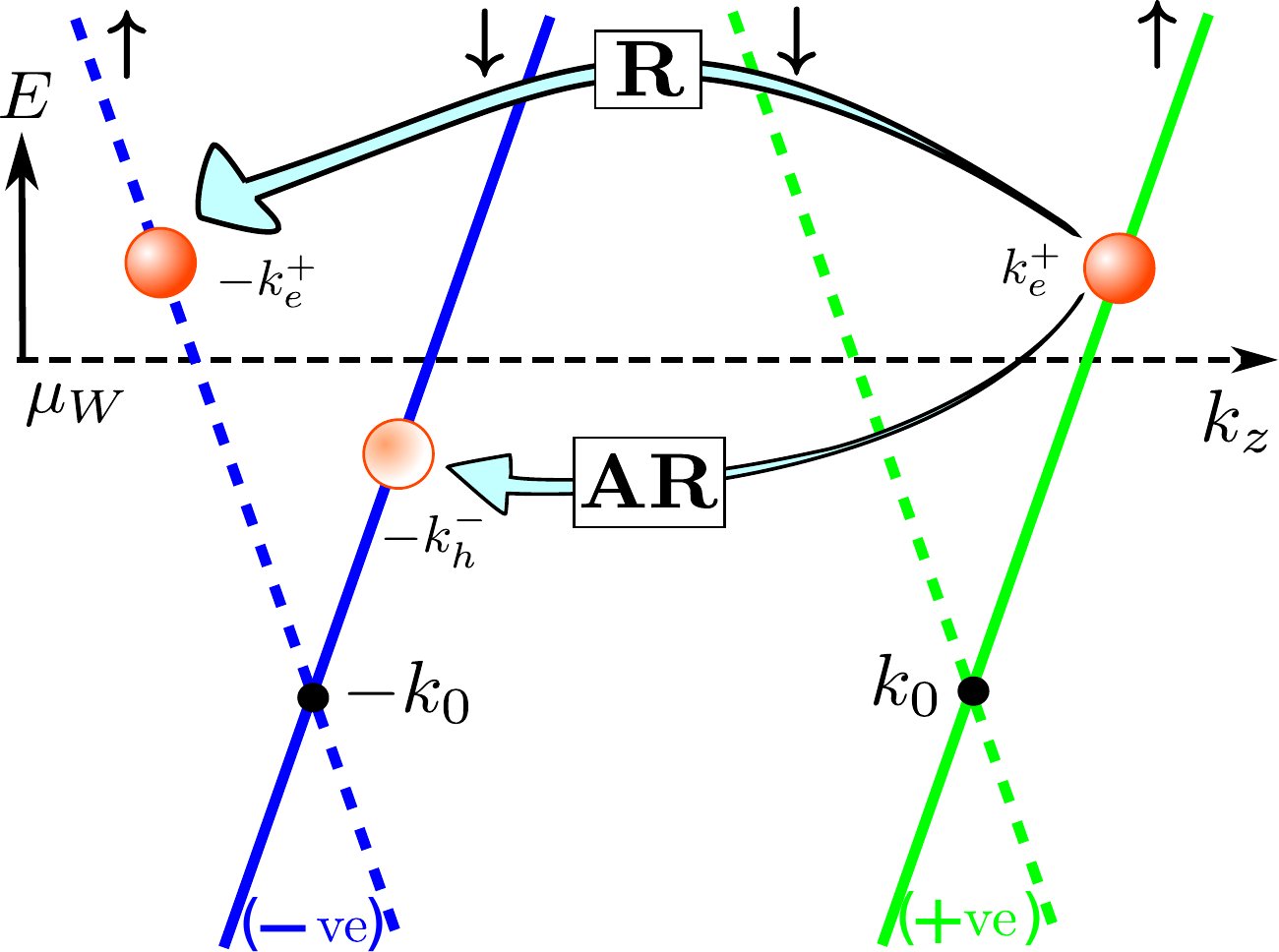}
  \caption{(Color online) Both reflection (R) and Andreev reflection (AR) in a WSM occur from one chiral node to another. The chiralities of the nodes are denoted as $+$ve and $-$ve, whereas the solid and the dashed lines show dispersions of the fermions at the two Weyl nodes with positive and negative velocities ($= d E/dk$). The  spins of the bands (denoted by arrows)  and the need to conserve spin
   accounts for the change of chirality for both normal and Andreev reflection.  }
   \label{randar}
\end{figure}

We have also studied\cite{josephson} the reflection and Andreev reflection  
(the phenomenon where an electron incident on a superconductor bounces back
as a hole and  two electrons (a Cooper pair) goes into the superconductor)
that takes place at the junction between a Weyl semi-metal and a superconductor.
A scattering approach\cite{uchida}  showed that the differential conductance depended
on the angle between the current and the vector  connecting the two Weyl points.
For the simplest model of a time-reversal breaking Weyl semimetal with two nodes,  we showed that both
normal and Andreev reflection  change chirality - see Fig.\ref{randar}\cite{josephson}.
We also showed that the existence of a new momentum scale introduced by the time-reversal
leads to the non-vanishing of normal reflection, unlike in graphene, where at low energies close
to Fermi energy, the current is purely due to Andreev reflection and normal reflection vanishes.

\begin{figure}
  \includegraphics[width=0.45\textwidth]{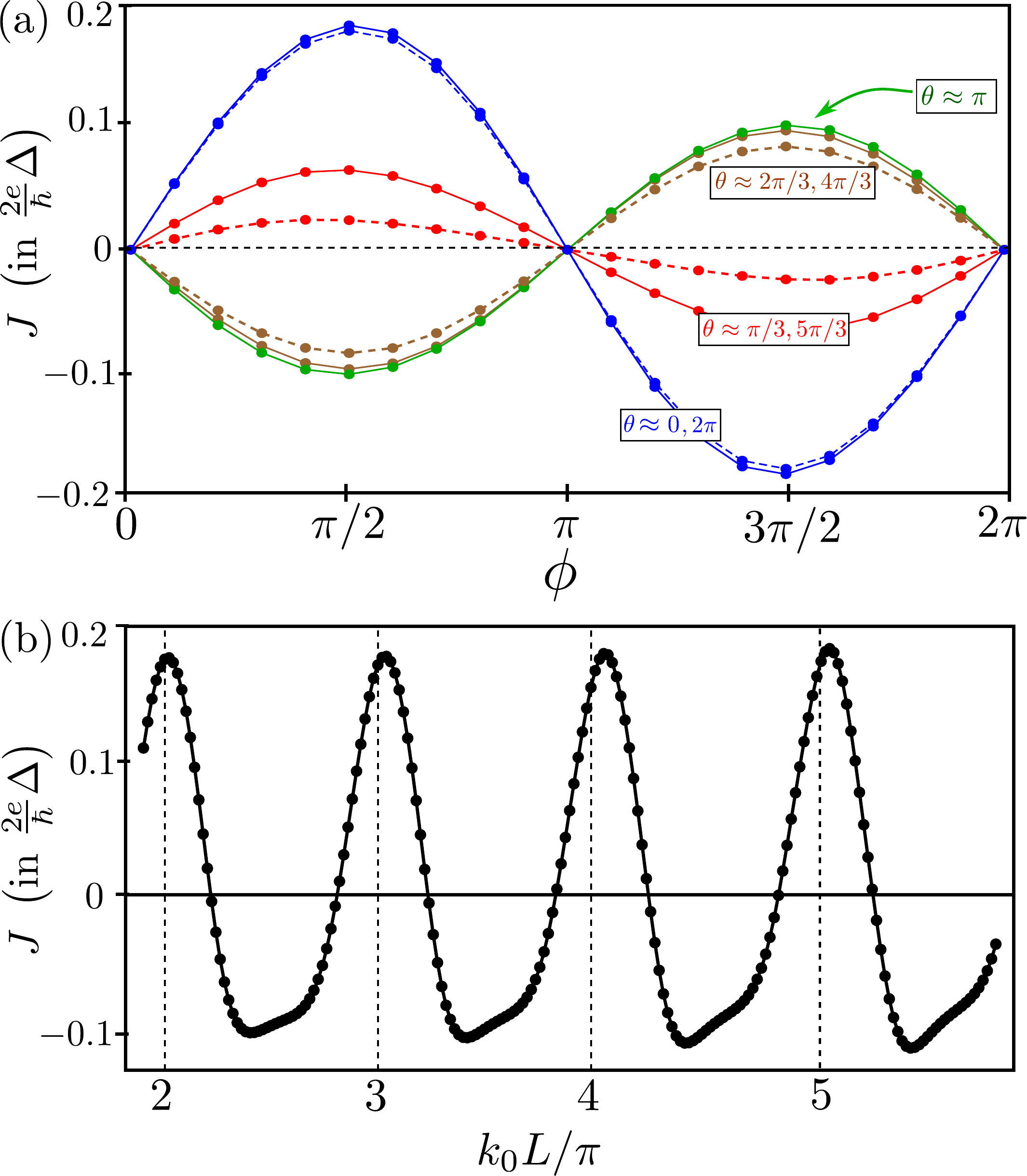}
  \caption{(Color online)  (a) Josephson current as a function of the superconducting phase difference $\phi$ for various values 
of $\theta=2k_0L ~\text{mod}(2\pi)$ and (b) Josephson current as a function of the length  ($k_0 L/\pi$) at $\phi=\pi/2$. The parameters used 
are  $\tilde{t}=0.25t$, $\epsilon=6t$, $\lambda=\lambda_z=t$, $\Delta=0.01t$, $\mu=0.02t$ and $L=60$ sites. For these parameters, the position of 
Weyl nodes, $k_0 \approx b_z$.}

   \label{jvsk0l}
\end{figure}

We also showed that when the Weyl semi-metal is sandwiched between two superconductors, the
Josephson current shows unexpected oscillations as a function of the time-reversal breaking
parameter  as shown in Fig.\ref{jvsk0l}. For normal metals, the current is expected to be independent
of the length $L$ for short ballistic Josephson junctions. For graphene, which is a Dirac metal, it was
shown that the critical current has diffusion-like scaling proportional to $1/L$ at the Dirac point, but 
without any impurity scattering\cite{titov}. But for the Weyl semi-metal, apart from the $2\pi$ periodicity
in $\phi$ (which is the phase difference between the two superconductors), we also find a periodicity in 
the length $L$ with period $\pi/k_0$. This is shown in Fig.\ref{jvsk0l}\cite{josephson}. Moreover, we also find the zero-pi transition\cite{zeropi}
of the Josephson current characteristic of superconductor-ferromagnet-superconductor junctions, which is perhaps
not surprising, since the WSM that we studied violated time-reversal invariance.
Here the Josephson current was computed
for the lattice model of the WSM studied in Sec.(V) through its Green\rq{}s function, $g (\omega) = \left[(\omega + i \delta) \mathcal{I} - H_0\right]^{-1}$, which was coupled to
two superconductors on either side of it, through an onsite self-energy\cite{proximity}
\begin{align}
 \Sigma_i (\omega) = \frac{\tilde{t}}{\sqrt{\Delta^2 - \omega^2}} (\mathcal{I}_{\tau} + \tau^{x}) [\omega \mathcal{I}_{\zeta} - \Delta e^{i \phi_i} \zeta^{x}] \mathcal{I}_{\sigma}.
\end{align}
Here $\zeta$ acted on the particle-hole degree of freedom of the model defined in the Nambu basis and 
$\Sigma_i$ was  defined only on the sites in contact with the $i^{\text{th}}$ superconductor. $\Delta$ denoted the
$s$-wave pair potential in the superconductor.
Then writing the full Green's function as $G (\omega) = (g^{-1}(\omega) - \Sigma_{L}(\omega) - \Sigma_{R}(\omega))^{-1}$,
 we computed the Josephson current\cite{josephson}.

\section{Future challenges and opportunities}

The experimental discovery of Weyl semi-metals may lead to new and unexpected spintronic applications. Weyl fermions
are gapless and hence have high mobility and can travel with very little resistance.They have topological protection because once a Weyl node is formed, it
can only annihilate if there is another Weyl node of opposte chirality and they are brought together. They are also 
 chiral and their spin is aligned to the direction of motion.

In the next few years, there is considerable scope in the experimental front for discovering new candidate materials for Weyl semi-metals,
particularly with time-reversal symmetry breaking, which has not yet been seen.  On the theoretical front, massless Weyl fermions
had been studied for many years in high energy physics as a possible candidate of neutrinos, before the mass for neutrinos was
discovered. It would be of interest to see whether any of the more exotic effects like neutrino oscillations have  possible manifestations
in the condensed matter context. The study and  classification of gapless topological materials is also far from complete and will
also probably gain centre-stage in the next few years.

\section*{Acknowledgments}  I would like to thank my collaborators Udit Khanna, Arijit Kundu and Dibyakanti Mukherjee for many useful
discussions on many aspects of topological insulators and Weyl semi-metals.

\end{document}